**Electrical transport and persistent photoconductivity in monolayer MoS$_2$ phototransistors**

Antonio Di Bartolomeo[1,2,*], Luca Genovese[1,3], Tobias Foller[3], Filippo Giubileo[2], Giuseppe Luongo[1,2], Luca Croin[4], Shi-Jun Liang[5], L. K. Ang[5], and Marika Schleberger[3]

[1]Dipartimento di Fisica "E. R. Caianiaello", Università di Salerno, via Giovanni Paolo II, Fisciano, 84084, Italy.
[2]CNR-SPIN Salerno, via Giovanni Paolo II, Fisciano, 84084, Italy
[3]Fakultät für Physik and CENIDE, Universität Duisburg-Essen, Lotharstrasse 1, 47057, Duisburg, Germany
[4]INRIM - Strada delle Cacce, 91 - 10135 Torino, Italy
[5]Engineering Product Development (EPD), Singapore University of Technology and Design (SUTD), 8 Somapah Road, 487372, Singapore

[*]Corresponding author. E-mail: adibartolomeo@unisa.it

**Abstract**

We study electrical transport properties in exfoliated molybdenum disulfide (MoS$_2$) back-gated field effect transistors at low drain bias and under different illumination intensities. It is found that photoconductive and photogating effect as well as space charge limited conduction can simultaneously occur. We point out that the photoconductivity increases logarithmically with the light intensity and can persist with a decay time longer than $10^4$ s, due to photo-charge trapping at the MoS$_2$/SiO$_2$ interface and in MoS$_2$ defects. The transfer characteristics present hysteresis that is enhanced by illumination. At low drain bias, the devices feature low contact resistance of $1.4\ k\Omega/\mu m$, ON current as high as $1.25\ nA/\mu m$, $10^5$ ON-OFF ratio, mobility of $\sim 1\ cm^2/Vs$ and photoresponsivity $\mathcal{R} \approx 1\ A/W$.

**Introduction**

Molybdenum disulfide (MoS$_2$) field effect transistors (FETs) have recently become very popular as alternatives to graphene devices for a new generation of transistors based on atomically-thin 2D materials [1-4]. Indeed, monolayer MoS$_2$ has a direct bandgap of $\sim 1.8 - 1.9\ eV$ [5], which enables the fabrication of transistors with ON/OFF ratio exceeding $10^8$ [6-7], that is suitable for logic applications and non-achievable with gapless graphene. Unfortunately, the success of MoS$_2$ based transistors has been hindered so far by the low mobility, which is in the range $0.01 - 50\ cm^2/Vs$ for MoS$_2$ on silicon dioxide (SiO$_2$) at room temperature [8]. Higher mobility has been achieved using MoS$_2$ together with high-k dielectrics and metal top-gate. The reduced Coulomb scattering due to the high-k dielectric environment and the subsequent possible modification of phonon dispersion in monolayer MoS$_2$ lead to mobilities of 100 $cm^2/Vs$ or higher [9, 10].

The direct bandgap of monolayer MoS$_2$ favors light absorption and enables efficient electron–hole pair generation under photoexcitation, making it an ideal layered material for optoelectronic applications. Both photoconductive and photogated transport, in which the channel conductance is enhanced by photogeneration or gating effect from trapped photocharge respectively, has been observed in MoS$_2$ phototransistors [11]. Ultrasensitive, monolayer MoS$_2$ phototransistors with internal gain and responsivity over a wide range, up to 2000 $AW^{-1}$ under visible light, have been demonstrated [12-17]. However, a strong variation in the switching time has been obtained, with photocurrent generation and annihilation times ranging from microseconds [17-19] to seconds [13-15]. For instance, Ref. [15] reports a slow photoresponse device operated at high drain voltage $V_{ds} = 8\ V$, characterized by rise and decay times of several seconds (~9 $s$). Slower photo-response has been demonstrated also in more recent works. Y.-C. Wu et al. [20] find a persistent photoconductivity effect, where current does not recover back to its original value even after a prolonged time period, and measure a decay time as long as $\tau \approx 5300\ s$ on monolayer MoS$_2$ back-gate transistors operated at a drain bias of 50 $mV$. The effect is attributed to random localized potential fluctuations in the devices, which originate from extrinsic sources based on the SiO$_2$/Si substrate. Persistent photocurrent has been also observed by K. Cho and coworkers [21] on FETs with multilayer MoS$_2$ nanosheets exposed to UV illumination under various environmental (vacuum or oxygen pressure) and measurement conditions. The observed photocurrent decay time exceeds $10^6\ s$, and faster decay occurs at higher oxygen pressure and gate voltage. Cho et al. attribute this result to charge trapping in oxygen related defect sites on the MoS$_2$ surface, and thus propose that trapping is affected by the application of the gate voltage. Attempts to reduces the persistent photoconductivity by suitable measurement conditions or passivation with high-k dielectric [19-20] or through the functionalization of MoS$_2$ [22-23] have been reported.

All the mentioned studies prove that the environmental and electrical measurement conditions affect the optoelectronic properties of MoS$_2$ FETs and that extrinsic or intrinsic charge trapping can play an important role in (photo)electronic transport. However, the physical mechanism for slow photoresponse of MoS$_2$, which seems to be an important peculiarity of this material, is still not fully understood. In particular, the origin of the observed persistent photoconductivity is still matter of debate.

In this paper, we fabricate back-gated transistors using cleaved monolayer MoS$_2$ to investigate electrical transport and photoresponse, and then uncover the cause of the slow photoresponse and persistent photoconductivity of MoS$_2$. To enhance the persistence effect, following ref [20], we focus on characterization at low drain bias. We measure gate- and light-dependence of FET electrical characteristics and a photogating effect, which clearly point toward charge trapping at the MoS₂/SiO₂ interface or in MoS₂ defects as the origin of the observed persistent photoconductivity in vacuum.

**Device fabrication**

Monolayer MoS₂ flakes were deposited on SiO₂/Si substrate by scotch-tape exfoliation of commercial crystalline bulk MoS₂ to fabricate back-gated FETs. The heavily doped p-type Si ($0.01 - 0.05\ \Omega cm$ resistivity) was used as the back gate with gate dielectric thickness $t_{SiO_2} = 285\ nm$. Metal leads, consisting of a Ti(5 nm)/Au(50 nm) bilayer, were evaporated after e-beam lithography to form source and drain in two- and four-probe configuration (Figure 1 (a)). Monolayer MoS₂ flakes were identified by optical microscopy and the number of layers was further confirmed by micro-Raman spectroscopy (Renishaw inVia, 532 nm wavelength). Figure 1 (b) shows the typical Raman spectrum of a selected monolayer MoS₂ flake used for device fabrication. The ~$19.7\ cm^{-1}$ separation between $E_{2g}^1$ and $A_{1g}$ Raman modes is a signature of monolayer MoS₂ [24].

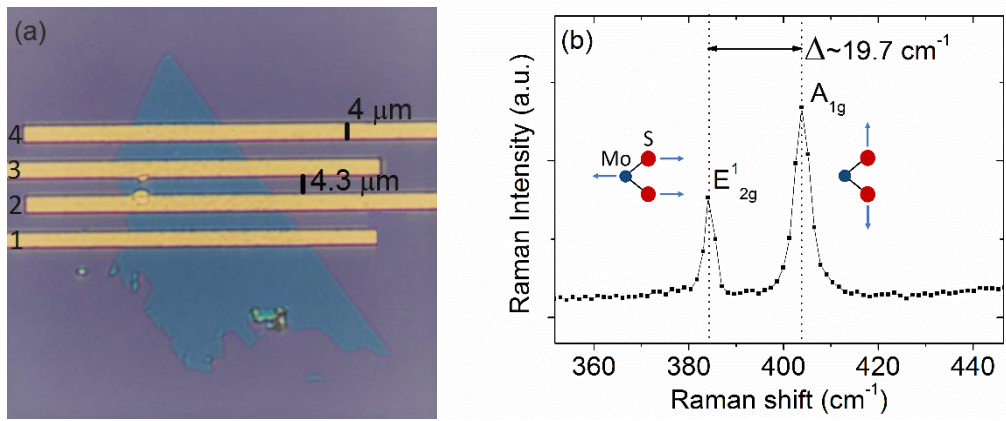

Figure 1 – (a) A MoS₂ flake on SiO₂/Si substrate with Ti/Au leads. (b) Raman spectrum of the MoS₂ flake shown in (a), with $E_{2g}^1$ and $A_{1g}$ peaks at ~$384.1\ cm^{-1}$ and ~$403.8\ cm^{-1}$, respectively.

Electrical measurements were performed in a Janis ST-500 cryogenic probe station connected to a Keithley 4200-SCS semiconductor parameter analyzer at room temperature and at an air pressure of 30 mbar to possibly remove moisture and limit oxygen adsorption on the MoS₂ surface. Optoelectronic properties were investigated by irradiation under visible light from an array of white LEDs (spectrum in the range 400-750 nm and peaks at 450 nm and 540 nm), with tunable intensity up to 5.5 $mW/cm^2$.

**Results and discussion**

Figure 2 (a) shows the transfer characteristic, $I_{ds} - V_g$ curve, in linear and logarithmic scale, obtained for the transistor of Figure 1 (a) using the inner leads 2 and 3 as the source and the drain, and the Si substrate as the back-gate, in dark and at air pressure of 30 mbar. The channel length $L$ and width $W$ are 4.3 $\mu m$ and 40 $\mu m$, respectively. The transistor shows n-type behavior with threshold voltage $V_T = -14\,V$ ($V_T$ is here defined as the gate voltage corresponding to a channel current $I_{ds} = 1nA$), an ON/OFF current ratio of $10^5$ and achieves ON current of $1.25\,nA/\mu m$ at $V_{ds} = 30\,mV$. Assuming $1/L$ scaling with channel length, the current drive is less than 2 orders of magnitude lower than that of state-of-the art Si FinFETs [25]. Figure 2 (b) shows linear output characteristics, $I_{ds} - V_{ds}$, measured in a two-probe configuration, which indicates good quality contacts, with ohmic behavior over the considered $V_{ds}$ range. Figure 2 (c) compares $I_{ds} - V_{ds}$ curves in a two- and four probe configuration, at grounded gate ($V_g = 0\,V$): From this figure, a contact resistance as low as $1.4\,k\Omega/\mu m$ is extracted, negligible with respect to the $M\Omega$ range of channel resistance. Such a result implies that the electrical measurements here presented are dominated by MoS$_2$ channel properties, rather than by contact resistance.

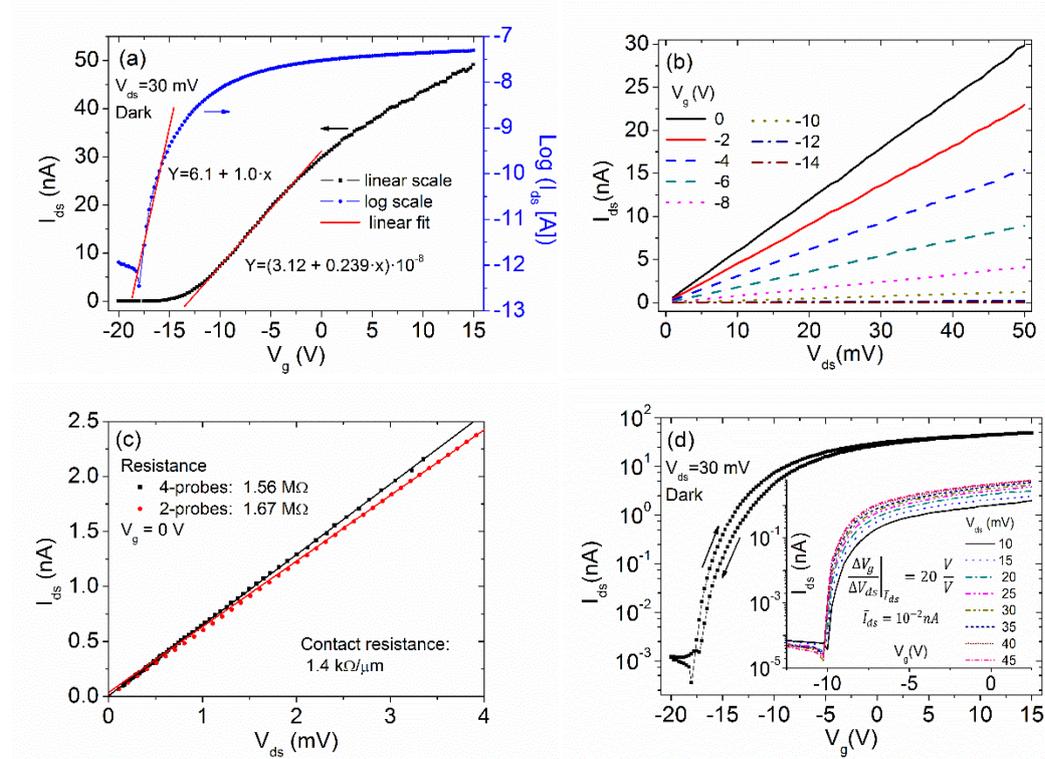

Figure 2 – Transfer (a) and output (b) characteristics measured between inner leads 2 and 3, in dark and at 30 mbar air pressure. (c) $I_{ds} - V_{ds}$ curve at $V_g = 0\,V$ measured in two and four-probe configuration to extract the contact resistance. (d) Hysteresis and (inset) effect of drain bias $V_{ds}$ on the transfer characteristic (the inset is referred to a different device from the same batch).

Figure 2 (a) shows a mobility

$$\mu = \frac{L}{W}\frac{1}{C_{SiO_2}}\frac{1}{V_{ds}}\frac{dI_{ds}}{dV_g} = \frac{L}{W}\frac{t_{SiO_2}}{\varepsilon_0\varepsilon_{SiO_2}}\frac{1}{V_{ds}}\frac{dI_{ds}}{dV_g} = 0.7\,\frac{cm^2}{Vs}$$

and a subthreshold slope

$$S = \left(\frac{dLogI_{ds}}{dV_g}\right)^{-1} = 2.3\frac{kT}{q}\left(1 + \frac{C_d}{C_{SiO_2}}\right) = \frac{1\,V}{decade}$$

The benchmark for S is the theoretical limit of $2.3\,kT/q \approx 60\,mV/decade$ obtained for the ideal metal-oxide semiconductor FET with $C_{SiO_2} \gg C_d$ (here, $q$ is the electron charge, $T$ the temperature, $k$ the Boltzmann constant, $C_{SiO_2} = 1.21 \cdot 10^{-8}\,F/cm^2$ is the SiO$_2$ capacitance per unit area and $C_d = C_{dep} + C_{tr}$ is the parallel of the depletion $C_{dep}$ and trap $C_{tr}$ capacitance per unit area). The obtained high value of S is due to the thick oxide which makes $C_{SiO_2} \ll C_d$. A high density of interface traps increases $C_d$ and can make the subthreshold slope worse [26].

Further transistor properties are shown in Figure 2 (d), where a hysteresis between forward and reverse sweep of $\Delta V_g = 1.2\,V$ at $\bar{I}_{ds} = 1\,nA$ and the effect of the increasing drain bias, resulting in augmented drive current and steeper subthreshold slope, are displayed. Such a hysteresis is quite common in unpassivated back-gated devices and is attributed to trap states and moisture either at the MoS$_2$/SiO$_2$ interface [27-30] or on the MoS$_2$ surface [31]. We can define a drain induced barrier lowering (DIBL) as $\left.\frac{\Delta V_g}{\Delta V_{ds}}\right|_{\bar{I}_{ds}} = 20\,\frac{V}{V}$ at $\bar{I}_{ds} = 10^{-2}\,nA$, which in this context represents the gate voltage change needed to achieve the same current produced by an increase of the drain bias of $\Delta V_{ds} = 1\,V$.

The quoted transistor figures of merit are consistent with the values reported in previous works for cleaved, air-exposed MoS$_2$ on SiO$_2$ [6, 20-21, 32-33].

To investigate the optoelectronic properties of the device, we checked the same parameters under illumination after keeping the device at the low air pressure of $30\,mbar$ for about 5 days.

Figure 3 (a) depicts the transfer and output characteristic under the white light of $5.5\,mW/cm^2$ of a LED array. Illumination keeps the linearity of the output characteristics (inset of Figure 3(a)), has negligible effect on channel mobility ($\mu = 0.6\,cm^2/Vs$), slightly increases the subthreshold slope ($S = 2V/decade$), but appreciably changes the threshold voltage ($V_T = -18\,V$) by shifting the transfer characteristic in the negative $V_g$ direction ($\Delta V_T \approx -4\,V$). Figure 3 (b) directly compares the transfer characteristics with light on and off, and shows that the OFF current is dramatically increased by more than two-orders of magnitude. Also, a hysteresis loop widening (to $\Delta V_g = 2.0\,V$ at $\bar{I}_{ds} = 1\,nA$) under illumination can be noticed.

The photocurrent, defined as the difference between device current with and without light,

$$I_{ds}^{Photo} = I_{ds}^{Light} - I_{ds}^{Dark},$$

is plotted in Figure 3 (c) and shows a strong dependence on the gate voltage with a maximum at $V_g \approx -10\,V$. The peak of photocurrent corresponds to the maximum photoresponsivity, defined as the ratio between the current and the total optical power $P_{opt}$ impinging on the device, that is

$$\mathcal{R} = \frac{I_{ds}^{Photo}}{P_{opt}} \approx 1\,\frac{A}{W}.$$

This value of $\mathcal{R}$ exceeds those quoted in ref. [12, 34] for similar devices, but is lower than the ones achieved in more recent works, with the devices treated or measured under special conditions [11,15].

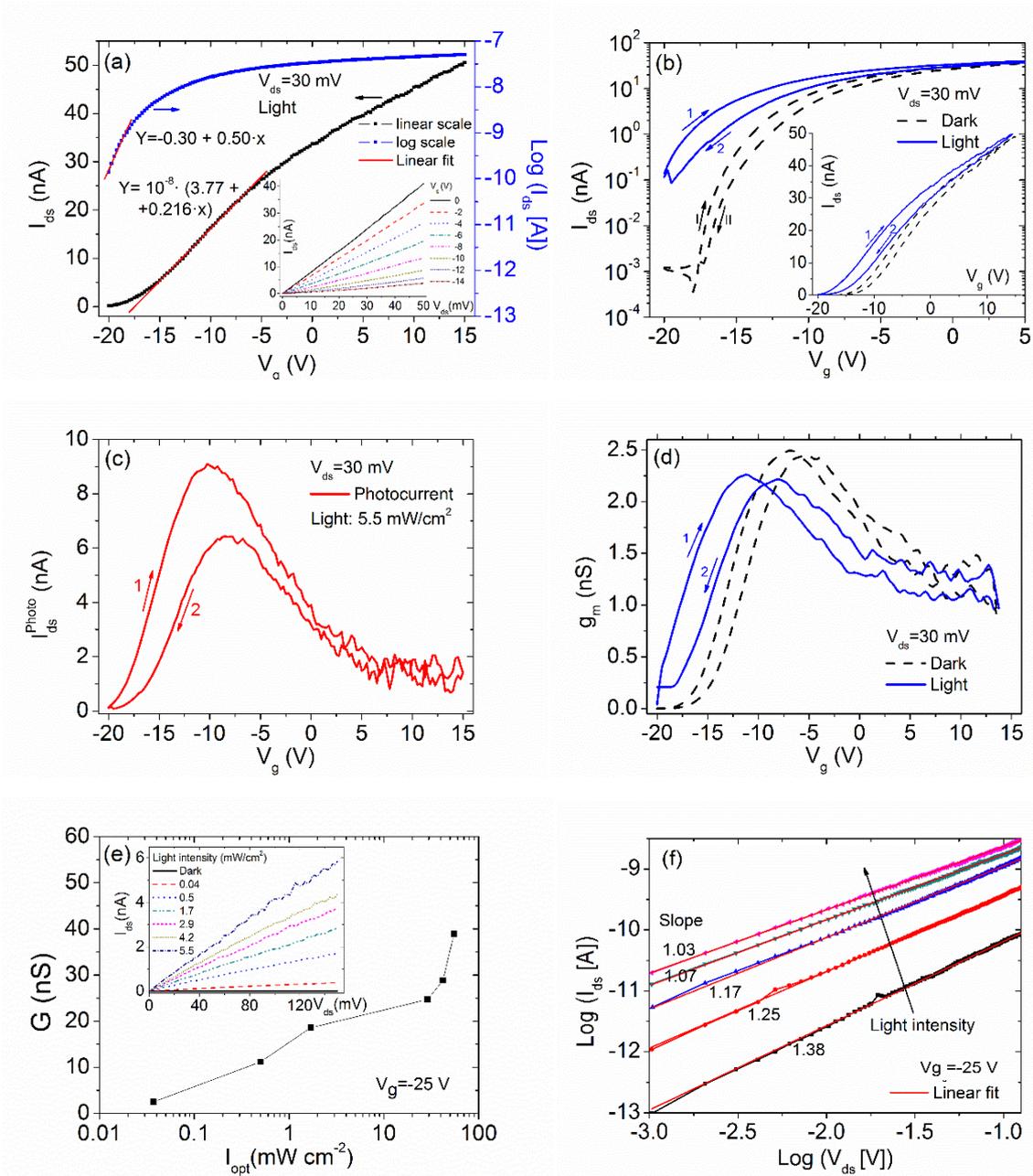

Figure 3 - Transfer and output (inset) characteristics under white LED light at $5.5\ mW/cm^2$. (b) Comparison of transfer characteristics in dark and under illumination (the inset shows the current in linear scale). (c) Photocurrent and (d) transconductance as a function of gate voltage. (e) Channel conductivity and output characteristics (inset) at $V_g = -25\ V$ as a function of light intensities. (f) Log-Log plot of $I_{ds}\ vs\ V_{ds}$ for different illumination levels.

In general, two different effects are assumed to occur in field effect phototransistors under illumination depending on the applied voltage $V_g$: The photogating (PG, often called photovoltaic) effect and the
66

photoconductive (PC) effect [35-36]. Hence, the photocurrent is the sum of two contributions: $I_{ds}^{Photo} = I_{ds}^{PG} + I_{ds}^{PC}$.

The PG effect manifests itself as a change in the transistor threshold voltage, from $V_T$ in the dark to $V_T + \Delta V_T$ under illumination, which gives a photocurrent [37-39] determined by

$$I_{ds}^{PG} = g_m \cdot |\Delta V_T| = \frac{A \cdot kT}{q} \ln\left(1 + \frac{\eta q P_{opt}}{I_{pd} h\nu}\right)$$

where $g_m = dI_{ds}/dV_g$ is the transistor transconductance, A is a fitting parameter, $\eta$ the quantum efficiency, $P_{opt}$ is the incident optical power, $I_{pd}$ is the dark current for holes, $h\nu$ is the photon energy. According to the above equation, the photocurrent due to photogating effect has the same $V_g$ dependence as the transconductance and scales logarithmically with $P_{opt}$. The photovoltaic effect originates from traps, which on optical excitation become charged, thus (photo)gating the device and shifting the threshold voltage [36]. The sign of $\Delta V_T$ indicates the polarity of the trapped carriers. Physically, the trapped carriers are located in defects of the channel itself or at the channel/oxide interface, where they act as an additional back gate.

The photoconductive effect refers to the increase in conductivity, $\Delta\sigma$, due to photogeneration of electron-hole pairs in the channel, and is less dependent on the gate voltage [35, 37]. Its contribution to the photocurrent is expressed as

$$I_{ds}^{PC} = \frac{W}{L} V_D \Delta\sigma.$$

Comparing Figure 3 (c) and 3 (d), we notice a strong similarity between the gate dependence of $I_{ds}^{Photo}$ and $g_m$. Both have peaks, which can be used to estimate $\Delta V_T$, using the equation for $I_{ds}^{PG}$. Remarkably, the shift found here of $\Delta V_T \approx -3.7\ V$ is compatible with the variation in $V_T$ estimated previously, using the transfer characteristics in dark and under illumination. Therefore, we conclude that the photoresponse is dominated by the PG effect, also because the PC effect is not expected to show an appreciable dependence on gate voltage. Note that the PC effect may be dominant in the region $V_g > 5V$ where the gate dependence of the current is smoother. Furthermore, Figure 3 (e) shows that the channel conductivity increases roughly with the logarithm of light intensity, implying again the presence of the PG effect.

All the electrical features discussed so far can be explained considering that photogenerated holes are trapped either in the uncompensated dangling bonds or in water molecules at the $SiO_2/MoS_2$ interface (extrinsic trapping) [40-42] or in the structural defects created in the $MoS_2$ [32, 42] (intrinsic trapping). Charge trapping in adsorbates as oxygen molecules on the top $MoS_2$ surface, proposed in previous works [31, 44], seems unlikely in this study because $O_2$ should favor negative charge trapping, while the leftwards shift of the transfer characteristic indicates that mainly positive charge is temporarily stored along the channel [45].

Noticeably, intrinsic or extrinsic traps sites are expected to provide localized and interface bandgap states, which on one hand can ease light absorption and on the other hand can introduce space charge limiting the conduction [32, 46]. To better understand the conduction mechanism in the channel, we plot $I_{ds} - V_{ds}$ for growing illumination in Log-Log scale as displayed in Figure 3 (f). The current follows a power law, $I_{ds} \propto V_{ds}^{\gamma}$ with $1.0 \leq \gamma \leq 1.4$ ($\gamma$ is the slope of the linear fit in Figure 3 (f)) with exponent increasing at lower illumination. At the lower illumination, the dominant mechanism seems to be a Child-Langmuir (CL) law (which predicts $I \propto V^{3/2}$) in which the current is space charge limited (SCL) [40, 47-49]. It is worth to mention that the CL law is derived for SCL conduction in vacuum (not for solid with or without traps). For a traditional solid with traps, the SCL conduction is normally described by the Mark-Helfrich (MH) law of $I \propto V^{l+1}$ with $l = T_c/T > 1$, where $T_c$ is a parameter characterizing the exponential distribution in energy of the traps. Based on the traditional SCL models for solids, the scaling of $I \propto V^{\gamma}$ is $\gamma \geq 2$. Thus the traditional SCL models in solids will not be able to explain $\gamma < 2$ obtained here and by other experiments [40]. This inconsistency between the model and the experiment may be resolved by a recent SCL model [50] for Dirac materials with a band gap (such as MoS$_2$) in which the power scaling will become $\gamma = 3/2$ similar to the CL law at the limit of ultra-relativistic regime. The sub-scaling of $\gamma < 3/2$ may be due to other quantum effects that remain to be confirmed. For example, the quantum CL law will have a scaling $1/2 \leq \gamma \leq 3/2$ [51]. In addition to the voltage scaling, the channel length scaling will be different [50] due to the nano-contact size of MoS$_2$ (like a thin film), and the total SCL current density will also be enhanced [52-53]. Note that the SCL regime gradually transforms into an ohmic one since the increased availability of photogenerated carriers can better screen localized trapped charge. Figure 3(f) also shows that the dependence of current on light intensity is weaker at high voltage than at lower voltage. This is likely due to space charge limited conduction which manifests more at higher drain bias, as reported in [40].

We finally checked the transient characteristics of the phototransistor by measuring the channel current under intermittent light, at given gate bias ($V_g = -25\ V$). The result is reported in Figure 4 (a). An abrupt, two-order of magnitude, increase in $I_{ds}$ is observed as soon as the light is turned on, as result of the photoexcitation (PC effect); this is followed by a much slower current increase, a behavior that can be understood considering that photogenerated holes get trapped and attract more electrons, which gradually increase the channel conductivity (PG effect). Switching off the light results in a very slow decay of the current. This is because trapped charges (PG effect) cause long sustained conductivity, despite the absence of illumination. The decay curve can be fitted using two exponentials

$I = I_0 + A_1 \exp(-t/t_1) + A_2 \exp(-t/t_2))$,

with shorter ($t_1 \sim 16\ s$) and longer ($t_2 \sim 350\ s$) time constants. The observed persistent photoconductivity is affected by the history of the device and can be enhanced by prolonged light exposure, as demonstrated in Figure 4 (b) where a further illumination of $500\ s$ increases the time decay constants up to $10^4\ s$.

Finally, Figure 4 (c) shows the time evolution of the transfer characteristic during the photoconductive state decay, corresponding to a gradual positive shift and to a small reduction of the hysteresis, which are indicative of slow positive charge detrapping.

We attribute the shorter decay component to photoconductive effect and to shallower traps, while the longer component is due to photogating effect caused by deeper (middle gap) traps, which are characterized by longer trapping/detrapping times (Figure 4 (d)). Indeed these traps require longer exposure to be filled, and yield the consequent very long component.

The origin of the trap states is still under debate. Persistent photoconductivity in $MoS_2$ devices has been already reported, and shown to be enhanced by charge trapping in the oxygen-related defect sites on the $MoS_2$ surface [21] or attributed to random local potential fluctuations which originate from extrinsic sources based on the substrate or defects in $MoS_2$ itself [20]. Our measurements convincingly point to the $MoS_2/SiO_2$ interface as the main charge trapping physical location. The absence of any hydrogen passivation of dangling bonds or annealing to desorb moisture at the $SiO_2$ surface prior to $MoS_2$ deposition obviously contribute to enhance the persistent photocurrent signature.

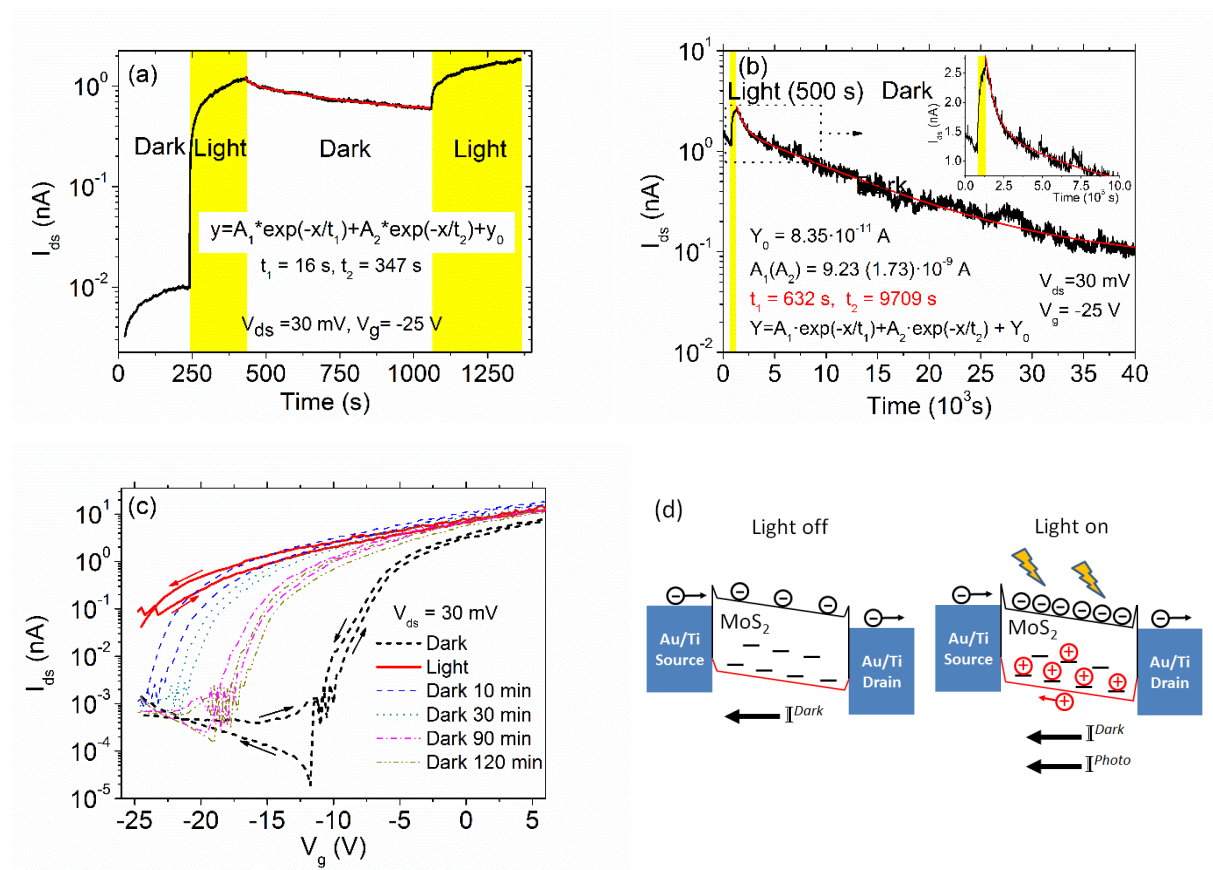

Figure 4 – Channel current versus time with and without light (a) and photocurrent decay after a longer light pulse (b). Time evolution of the transistor transfer characteristic after a light pulse (c). The light intensity was set to 5.5 $mW/cm^2$. (d) Band diagram of the device without and with light, showing hole trapping in shallower and deeper trap states close to the valence band. Hole trapping causes a down-shift of $MoS_2$ bands and an increase of the electron density, that is of the channel conductivity.

## Conclusion

In summary, we studied electrical transport in exfoliated $MoS_2$ field effect transistors at low drain bias and under different illumination intensities. The transistors exhibited n-type behavior with threshold voltage depending on the illumination level. Light also affects the hysteresis observed in transfer characteristics. The long persistent photocurrent measured in our devices is most likely due to photo-charge trapping in $MoS_2$ defects and at the $MoS_2/SiO_2$ interface, which cause photogating effect. Apart from photogating and photoconductive effects, we showed that space charge limited conduction also appear at low illumination and possibly at higher drain bias.

## Acknowledgments

T. Foller would like to thank the group of A. Lorke for access to the e-beam lithography system.

## References

[1] Butler S Z, Hollen S M, Cao L, Cui Y, Gupta J A, Gutiérrez H R and Johnston-Halperin E 2013 *ACS Nano* **7** 2898-2926

[2] Tong X, Ashalley E, Lin F, Li H and Wang Z M 2015 *Nano-Micro Letters* **7** 203-218

[3] Di Bartolomeo A, Giubileo F, Romeo F, Sabatino P, Carapella G, Iemmo L, Schroeder T and Lupina G 2015 *Nanotechnology* **26** 475202

[4] Di Bartolomeo A, Giubileo F, Iemmo L, Romeo F, Russo S, Unal S, Passacantando M, Grossi V, and A. M. Cucolo 2016 *Appl. Phys. Lett.* **109** 023510

[5] Mak K F, Lee C, Hone J, Shan J and Heinz T F. 2010 *Physical Review Letters* **105** 136805

[6] Radisavljevic B, Radenovic A, Brivio J, Giacometti I V and Kis A 2011 *Nature Nanotechnology* **6** 147-150

[7] Wu W, De D, Chang S C, Wang Y, Peng H, Bao J and Pei S S 2013 *Applied Physics Letters* **102** 142106

[8] Fuhrer M S and Hone J 2013 *Nature Nanotechnology* **8** 146-147

[9] Radisavljevic B and Kis A 2013 *Nature Materials* **12** 815-820

[10] Wang J, Yao Q, Huang C W, Zou X, Liao L, Chen S and Jiang C 2016 *Advanced Materials* **28** 8302-8308

[11] Furchi M M, Polyushkin D K, Pospischil A and Mueller T 2014 *Nano Letters* **14** 6165-6170

[12] Yin Z, Li H, Li H, Jiang L, Shi Y, Sun Y and Zhang H 2011 *ACS Nano* **6** 74-80

[13] Lee H S, Min S W, Chang Y G, Park M K, Nam T, Kim H and Im S 2012 *Nano Letters* **12** 3695-3700

[14] Zhang W, Huang J K, Chen C H, Chang Y H, Cheng Y J and Li L J 2013 *Advanced Materials* **25** 3456-3461

[15] Lopez-Sanchez O, Lembke D, Kayci M, Radenovic A and Kis A 2013 *Nature Nanotechnology* **8** 497-501

[16] Fontana M, Deppe T, Boyd A K, Rinzan M, Liu A Y 2013 *Scientific Reports* **3** 1634

[17] Cao B, Shen X, Shang J, Cong C, Yang W, Eginligil M and Yu T 2014 *APL Materials* **2** 116101

[18] Tsai D S, Liu K K, Lien D H, Tsai M L, Kang C F, Lin C A and He J H 2013 *ACS Nano* **7** 3905-3911

[19] Kufer D and Konstantatos G 2015 *Nano Letters* **15** 7307-7313

[20] Wu Y C, Liu C H, Chen S Y, Shih F Y, Ho P H, Chen C W and Wang W H 2015 *Scientific Reports* **5** 11472

[21] Cho K, Kim T Y, Park W, Park J, Kim D, Jang J and Lee T 2014 *Nanotechnology* **25** 155201

[22] Rathi N, Rathi S, Lee I, Wang J, Kang M, Lim D and Kim G H 2016 *RSC Advances* **6** 23961-23967

[23] Pak J, Jang J, Cho K, Kim T Y, Kim J K, Song Y and Lee T 2015 *Nanoscale* **7** 18780-18788

[24] Li H, Zhang Q, Yap C C R, Tay B K, Edwin, T H T, Olivier A and Baillargeat D 2012 *Advanced Functional Materials* **22** 1385-1390

[25] Natarajan S, Agostinelli M, Akbar S, Bost M, Bowonder A, Chikarmane V and Ghani T A *2014 IEEE International Electron Devices Meeting (IEDM)*, (pp. 3-7).

[26] Ayari A, Cobas E, Ogundadegbe O and Fuhrer M S 2007 *Journal of Applied Physics* **101** 014507

[27] Park Y, Baac H W, Heo J and Yoo G 2016 *Applied Physics Letters* **108** 083102

[28] Di Bartolomeo A, Rinzan M, Boyd AK, Yang Y, Guadagno L, Giubileo F, and Barbara P 2010 *Nanotechnology* **21** 115204

[29] Di Bartolomeo A, Giubileo F, Santandrea S, Romeo F, Citro R, Schroeder T, Lupina G 2010 *Nanotechnology* **22** 275702

[30] Di Bartolomeo A, Santandrea S, Giubileo F, Romeo F, Petrosino M, Citro R, Barbara P, Lupina G, Schroeder T, Rubino A 2013 *Diamond and Related Materials* **38** 19–23

[31] Late D J, Liu B, Matte H R, Dravid V P and Rao C N R 2012 *ACS Nano* **6** 5635-5641

[32] Qiu H, Xu T, Wang Z, Ren W, Nan H, Ni Z and Long G 2013 *Nature Communications* **4** 2642

[33] Newaz A K M, Prasai D, Ziegler J I, Caudel D, Robinson S, Haglund Jr R F and Bolotin K I 2013 *Solid State Communications* **155** 49-52

[34] Choi W, Cho M Y, Konar A, Lee J H, Cha G B, Hong S C and Kim S 2012 *Advanced Materials* **24** 5832-5836

[35] Saragi T P, Londenberg J and Salbeck J 2007 *Journal of Applied Physics* **102** 046104

[36] Buscema M, Island J O, Groenendijk D J, Blanter S I, Steele G A, van der Zant H S and Castellanos-Gomez A 2015 *Chemical Society Reviews* **44** 3691-3718

[37] Takanashi Y, Takahata K and Muramoto Y 1999 *IEEE Transactions on Electron Devices* **46** 2271-2277

[38] Choi C S, Kang H S, Choi W Y, Kim H J, Choi W J, Kim D H and Seo K S 2003 *IEEE Photonics Technology Letters* **15** 846-848

[39] Kang H S, Choi C S, Choi W Y, Kim D H and Seo K S 2004 *Applied Physics Letters* **84** 3780-3782

[40] Ghatak S and Ghosh A 2013 *Applied Physics Letters* **103** 122103

[41] Guo Y, Wei X, Shu J, Liu B, Yin J, Guan C and Chen Q 2015 *Applied Physics Letters* **106** 103109

[42] Illarionov Y Y, Rzepa G, Waltl M, Knobloch T, Grill A, Furchi M M and Grasser T 2016 *2D Materials* **3** 035004

[43] Lin Z, Carvalho B R, Kahn E, Lv R, Rao R, 2016 *2D Materials* **3** 022002

[44] Cho K, Park W, Park J, Jeong H, Jang J, Kim T Y and Lee T 2013 *ACS Nano* **7** 7751-7758

[45]Ochedowski O, Marinov K, Scheuschner N, Poloczek A, Bussmann BK, Maultzsch J, Schleberger 2014 *Beilstein J Nanotechnol.* **5** 291

[46] Ghatak S, Pal A N and Ghosh A 2011 *ACS Nano* **5** 7707-7712

[47] Mark P and Helfrich W 1962 *Journal of Applied Physics* **33** 205-215

[48] Kumar V, Jain S C, Kapoor A K, Poortmans J and Mertens R 2003 *Journal of Applied physics* **94** 1283-1285

[49] Nath C and Kumar A 2012 *Journal of Applied Physics* **112** 093704

[50] Ang Y S, Zubair M, and Ang L K 2017 *Physical Review B* (to be published)

[51] Ang L K, Kwan T J T and Lau Y Y 2003 *Physical Review Letters* **91** 208303

[52] Chandra W, Ang L K, Pey K L and Ng C M 2007 *Applied Physics Letters* **90** 153505

[53] Zhu Y B and Ang L K 2015 *Scientific Reports* **5** 9173